\documentclass[manuscript]{acmart}

\renewcommand\footnotetextcopyrightpermission[1]{}

\setcopyright{none}
\makeatletter
\def\@copyrightpermission{}
\makeatother

\AtBeginDocument{%
  }

\usepackage{array}
\DeclareUnicodeCharacter{202A}{}
\DeclareUnicodeCharacter{202C}{}
\DeclareUnicodeCharacter{202D}{}
\usepackage[utf8]{inputenc}
\usepackage[T1]{fontenc}
\usepackage{lmodern}
\makeatletter
\def\ACM@cc@type{}
\makeatother
\begin{document}

%%\title{Doing Audits Right? The Role of Sampling and Legal Content Analysis in Systemic Risk Assessments and Independent Audits in the Digital Services Act}
\title[Doing Audits Right?]{Doing Audits Right? The Role of Sampling and Legal Content Analysis in Systemic Risk Assessments and Independent Audits in the Digital Services Act}
%%\author{Ben Trovato}
%%\authornote{Both authors contributed equally to this research.}
%%\email{trovato@corporation.com}
%%\orcid{1234-5678-9012}
%%\author{G.K.M. Tobin}
%%\authornotemark[1]
%%\email{webmaster@marysville-ohio.com}
%%\affiliation{%
%%  \institution{Institute for Clarity in Documentation}
 %% \streetaddress{P.O. Box 1212}
 %% \city{Dublin}
 %% \state{Ohio}
 %% \country{USA}
 %% \postcode{43017-6221}
%%}

\author{Marie-Therese Sekwenz}
\affiliation{%
 \institution{Delft University of
Technology}
%%  \streetaddress{1 Th{\o}rv{\"a}ld Circle}
  \city{Delft}
 \country{Netherlands}}
\email{m.t.sekwenz@tudelft.nl}

\author{Rita Gsenger}
\affiliation{%
  \institution{Weizenbaum Institute}
  \city{Berlin}
  \country{Germany}
}
\email{rita.gsenger@weizenbaum-institut.de}

\author{Scott Dahlgren}
\affiliation{%
 \institution{Erasmus MC}
%%  \streetaddress{1 Th{\o}rv{\"a}ld Circle}
  \city{Rotterdam}
 \country{Netherlands}}
\email{fdahlgr@gmail.com}

\author{Ben Wagner}
\affiliation{%
 \institution{I:TU Interdisciplinary Transformation
University Austria, Delft University of
Technology, Inholland}
%%  \streetaddress{1 Th{\o}rv{\"a}ld Circle}
  \city{Linz, Delft}
 \country{Austria, Netherlands}}
\email{B.Wagner@tudelft.nl}

%%\author{Valerie B\'eranger}
%%\affiliation{%
%%  \institution{Inria Paris-Rocquencourt}
 %% \city{Rocquencourt}
%%  \country{France}
%%}

%%\author{Aparna Patel}
%%\affiliation{%
%% \institution{Rajiv Gandhi University}
%% \streetaddress{Rono-Hills}
%% \city{Doimukh}
 %%\state{Arunachal Pradesh}
%% \country{India}}

%%\author{Huifen Chan}
%%\affiliation{%
%%  \institution{Tsinghua University}
%%  \streetaddress{30 Shuangqing Rd}
%%  \city{Haidian Qu}
%%  \state{Beijing Shi}
%%  \country{China}}

%%\author{Charles Palmer}
%%\affiliation{%
 %% \institution{Palmer Research Laboratories}
 %% \streetaddress{8600 Datapoint Drive}
 %% \city{San Antonio}
 %% \state{Texas}
 %% \country{USA}
 %% \postcode{78229}}
%%\email{cpalmer@prl.com}

%%\author{John Smith}
%%\affiliation{%
 %% \institution{The Th{\o}rv{\"a}ld Group}
%%  \streetaddress{1 Th{\o}rv{\"a}ld Circle}
 %% \city{Hekla}
 %% \country{Iceland}}
%%\email{jsmith@affiliation.org}

%%\author{Julius P. Kumquat}
%%\affiliation{%
 %% \institution{The Kumquat Consortium}
%%  \city{New York}
 %% \country{USA}}
%%\email{jpkumquat@consortium.net}


\begin{abstract}
A central part of the European Union’s Digital Services Act (DSA) is the requirement for internal and external audits of online platforms.
A key component of these audits is the analysis of platform systemic risk assessments, which cover wide-ranging and multifaceted
topics such as the dissemination of illegal content, negative effects on human rights, the electoral process, and gender-based violence.
The Delegated Regulation on how to conduct such audits under the DSA further specifies what is expected from platforms and auditors
to ensure compliance with the law. This article discusses the pros and cons of different quantitative and qualitative methods that
could be used to audit the varying systemic risks of online platforms and search engines under Europe’s DSA regime. Following an
evaluation of key methods, we propose an empirical mixed-method approach based on risk-specific sampling techniques combined
with (legal) content analysis as a reliable auditing method for the DSA. We argue that audits should be based on the evaluation of
content samples representing spot checks for systemic risks. To develop this mixed-method approach, we first analyse the European
provisions relevant to auditable content samples and their selection. Then, we explore sampling techniques that can be meaningfully
applied in internal and external DSA audits and assess methodological techniques for testing systemic risk categories. Furthermore, we
examine how the representativeness of a sample can be understood in different disciplines. Finally, we analyse the first systemic risk
assessment reports of platforms with a view to understanding how the tests and methodologies (such as sampling techniques) used in these assessments evaluate systemic
risks. By proposing a novel empirical mixed-method approach for conducting DSA audits, this article addresses the complex challenges
of evidence-based risk assessment in content moderation. Furthermore, we contribute to the rapidly growing field of DSA compliance
assessments, emphasizing the specificities of audits and systemic risks. In doing so, we recognize that there is no “one-size-fits-all”	solution for DSA audits, but rather a use case-specific and platform-tailored approach is needed.

\end{abstract}

%%\begin{CCSXML}
%%<ccs2012>
%% <concept>
%%  <concept_id>00000000.0000000.0000000</concept_id>
%%  <concept_desc>Do Not Use This Code, Generate the Correct Terms for Your Paper</concept_desc>
%%  <concept_significance>500</concept_significance>
%% </concept>
%% <concept>
%%  <concept_id>00000000.00000000.00000000</concept_id>
%%  <concept_desc>Do Not Use This Code, Generate the Correct Terms for Your Paper</concept_desc>
%%  <concept_significance>300</concept_significance>
%% </concept>
%% <concept>
%%  <concept_id>00000000.00000000.00000000</concept_id>
%%  <concept_desc>Do Not Use This Code, Generate the Correct Terms for Your Paper</concept_desc>
%%  <concept_significance>100</concept_significance>
%% </concept>
%% <concept>
%%  <concept_id>00000000.00000000.00000000</concept_id>
%%  <concept_desc>Do Not Use This Code, Generate the Correct Terms for Your Paper</concept_desc>
 %% <concept_significance>100</concept_significance>
%% </concept>
%%</ccs2012>
%%\end{CCSXML}
\begin{CCSXML}
<ccs2012>
   <concept>
       <concept_id>10003456.10003462</concept_id>
       <concept_desc>Social and professional topics~Computing / technology policy</concept_desc>
       <concept_significance>500</concept_significance>
       </concept>
  
\end{CCSXML}

\ccsdesc[500]{Social and professional topics~Computing / technology policy}

%%\ccsdesc[500]{Do Not Use This Code~Generate the Correct Terms for Your Paper}
%%\ccsdesc[300]{Do Not Use This Code~Generate the Correct Terms for Your Paper}
%%\ccsdesc{Do Not Use This Code~Generate the Correct Terms for Your Paper}
%%\ccsdesc[100]{Do Not Use This Code~Generate the Correct Terms for Your Paper}

\keywords{Digital Services Act, Sampling, Systemic Risk Assessments, Auditing}

%%  \includegraphics[width=\textwidth]{sampleteaser}
%%  \caption{Seattle Mariners at Spring Training, 2010.}
%%  \Description{Enjoying the baseball game from the third-base
%%  seats. Ichiro Suzuki preparing to bat.}

%%\received{20 February 2007}
%%\received[revised]{12 March 2009}
%%\received[accepted]{5 June 2009}

\maketitle

\section{Introduction}
The Digital Services Act (DSA) \cite{RegulationEU20222022b} represents a cornerstone of Europe’s efforts to regulate online expression and intermediary services and aims to harmonize rules for user-generated and synthetic content on online platforms. The DSA aims to increase transparency and protect users from illegal content online. It aspires to achieve the so-called ``Brussels effect'', potentially influencing global regulatory standards  \cite{nunziatoDigitalServicesAct2023}. It entered into force on February 17, 2024.
The DSA applies to \emph{intermediary services} operating within the European Union (EU) according to Art. 2(1). These
services offer network infrastructure, such as hosting platforms (for example, cloud services) or online platforms
(such as app stores or online marketplaces). Importantly, these services also include Very Large Online Platforms (VLOPs) and
Very Large Search Engines (VLOSEs) \cite{RegulationEU20222022b}, which face the most stringent requirements. They are defined
as platforms with over 45 million average monthly users (or 10 \% of the EU population) in the European Union (Art. 33 DSA). The European Commission published (and continuously updates) a list of designated VLOPs and VLOSes, including platforms such as Amazon, LinkedIn, Wikimedia, and Facebook 
\cite{commission_supervision_2024}.

Therefore, the Regulation adopts a layered approach to obligations based on platform size and function, considering an increased risk for society due to content that is accessible to a large number of users. The DSA includes various new obligations for users, businesses, and platforms. Key provisions include the adaptation of Terms and Conditions (TaC) (Art. 14), transparency obligations (Arts. 15, 24, 42), the design of notice and take-down mechanisms (Art. 16), platforms’ content moderation decision reporting (Art. 17), Trusted Flagger involvement (Art. 22), recommendation system disclosures (Art. 27), researcher access for independent evaluations (Art. 40), ex ante risk assessments (Arts. 34, 35), or third-party audits (Art. 37). These external audits are done by auditing firms like KPMG, EY, or Deloitte \cite{noauthor_dsa_nodate}.
Recent enforcement actions underscore the DSA’s impact. For example, on December 18, 2023, the European Commission initiated formal proceedings against X (formerly Twitter) for noncompliance, citing risks related to illegal content dissemination (Art. 34(1)(a)) and threats to civic discourse and electoral processes (Art. 34(1)(c))  \cite{CommissionOpensFormal}. The Commission also opened proceedings against TikTok on election risks \cite{noauthor_commission_2024} and against Temu for the sale of illegal products, their addictive platform design, failed disclosure of recommendation system, and failed compliance of researcher data access  \cite{CommissionTemu2024}.

\section{Background}

Various research endeavours have already looked into audits in the online space in general and the auditing requirements of the DSA specifically. Research has focused on auditing algorithms \cite{imana_having_2023, terzis_law_2024}  and their challenges. More recently, AI audits have emerged as a discussion point, as increasingly, approaches are proposed to audit AI systems\cite{lam_framework_2024, falco_governing_2021} to increase fairness and accountability\cite{calvi_enhancing_2023}. Furthermore, some research elaborates on increasing auditability, for instance, in extended reality (XR) applications \cite{norval_navigating_2023, gray_legal_2024}. These studies have highlighted issues with accessing systems, highlighting that black-box access is not enough \cite{casper_black-box_2024}.  AI can be used to perform audits  \cite{li_making_2024}; however, these audits can be biased \cite{groves_auditing_2024}. European Regulations are not the first to establish auditing as a mechanism of transparency. Moreover, the DSA, the Digital Markets Acts \cite{RegulationEU20222022a}, and the AI Act  \cite{RegulationEU20242024}   include auditing provisions. Effective auditing under the DSA requires a robust framework that ensures auditor independence and competence. Kaushal et al. propose guidelines for audit quality assurance, including performance standards and quality control procedures, which are essential for credible assessments\cite{kaushal_automated_2024}. They provide an empirical analysis of the DSA’s transparency database, highlighting the legal obligations and the role of automated systems in ensuring compliance. Terzis et al. highlight that although the auditing mechanisms are not new, they have a larger scope than other legally required audits. However, they only apply to a smaller number of audited entities (namely, to VLOPs and VLOSEs) \cite{terzis_law_2024}. Moreover, to align sampling with specific risks, such as the dissemination of illegal content (Art. 34(1)(a) DSA), auditors must ensure that samples represent the population accurately, particularly in ex post content moderation scenarios. Studies such as Mubarak et al.’s work on disinformation emphasize the value of comparing deleted and undeleted content samples \cite{mubarakDetectingReasoningDeleted2023}.
Various challenges in auditing remain, and as such, ``auditing the auditors'', finding best practices, and standards are still necessary  \cite{costanza-chock_who_2022}. Additionally, no public participation or involvement of data subjects is intended, even though data access for vetted researchers, as foreseen by Art. 40 DSA, is a first step towards some sort of public involvement   \cite{calvi_enhancing_2023}. Finally, the involvement of traditional auditors in algorithmic audits introduces complexities, such as potential conflicts of interest and the need for specialized expertise, which must be addressed to ensure effective oversight\cite{terzis_law_2024}.  
Per Art 34 and 37 DSA, audits assess content moderation quality, addressing risks such as illegal content the assessment of content moderation quality in detail to understand if, for example, the risk of disseminating illegal content is adequately addressed. This process, however, is criticized for being opaque and arbitrary 
\cite{gomez_algorithmic_2024}.
As a result of  the volume and diversity of online content \cite{DomoResourceData}, challenges of varied risk types  \cite{aliUnderstandingEffectDeplatforming2021,binnsTrainerBotInheritance2017} and dynamic time constraints arise. The Delegated Regulation (DR) accompanying the DSA provides additional guidance for conducting audits, emphasizing the need for transparent frameworks and methodologies\cite{DelegatedRegulationSupplementing2023}. It clarifies the need for frameworks and methodologies used in the process of assessing risks and auditing, including a justification for why specific methods were chosen. These methodologies should include tests to assess compliance with the DSA (Art. 1(c) DR). Sampling methods, explicitly mentioned in Art. 10 (DR), are integral to these evaluations and widely discussed in the academic literature\cite{bilonEvaluationSamplingMethods2020a, brightStatisticalSamplingTax1988, davidsonAccuracyStatisticalSampling1959, pina-garciaStandardSamplingMethodology2016}. This paper focuses on sampling methodologies under the DSA, addressing two key research questions:
(RQ1) What sampling methods are most relevant for assessing compliance under the DSA for internal risk assessments and independent audits?, 
%\item Which sampling techniques best fit the systemic risk categories defined in Art 34 DSA? (RQ2)
and (RQ2) How do systemic risk assessment reports and audit reports from VLOPs and VLOSEs address the use of sampling methodologies, and what are the implications of their variability for the standardization of audit practices? (RQ3)
The following sections provide an in-depth exploration of these questions.

\section{AUDITING VERY LARGE ONLINE PLATFORMS AND VERY LARGE SEARCH ENGINES IN THE DIGITAL SERVICES ACT}

The DSA establishes a rigorous framework for auditing VLOPs and VLOSEs; however, in several aspects, the framework is not unambiguous  \cite{terzis_law_2024}. These auditing processes have two central components: internal systemic risk assessments (Art. 34 DSA) and external independent audits (Art. 37 DSA). These processes culminate in the generation of key deliverables, including risk assessment reports prepared by platforms (Art. 42(4)(a) DSA; Rec. 100) and audit reports authored by independent auditors (Art. 37(4) DSA; Art. 6 DR). If an audit report identifies deficiencies, platforms must implement corrective measures, document the process in an audit implementation report, and undergo re-evaluation (Art. 37(6) DSA). Systemic risk assessments are mandated annually under Art. 34(1) DSA and address four critical categories: (a) dissemination of illegal content, (b) adverse impacts on fundamental rights, (c) negative effects on electoral processes, threats to civic discourse and public security, and (d) harms related to gender-based violence, public health, and individual well-being. Platforms must evaluate their algorithmic systems (Art. 34(2)(a)), content moderation mechanisms (Art. 34(2)(b)), enforcement of Terms and Conditions (Art. 34(2)(c)), advertising systems (Art. 34(2)(d)), and data-related practices (Art. 34(2)(e)) to address these risks effectively. This layered approach reflects broader discussions in the fairness literature in the FAccT community about designing contextually aware and compliant systems. For example, Laufer et al. highlight the need for interdisciplinary collaboration to effectively address challenges like systemic risks, a principle that underpins the DSA’s framework for compliance and auditing  \cite{laufer_four_2022}.
Platforms must also implement mitigation measures for identified risks (Art. 35 DSA). Independent audits, as outlined in Art. 37 DSA, are tasked with verifying compliance across a comprehensive range of requirements outlined in Arts. 10–48 of the DSA. These requirements reflect a critical shift from merely enforcing compliance to addressing the broader socio-technical dynamics that influence systemic risks, as Rateike et al. explored in their discussion on designing long-term fair policies in dynamic systems \cite{rateike_designing_2024}. The audit process is integral to the DSA’s enforcement strategy, ensuring that platforms  not only comply with existing regulations but also actively address systemic risks. A ``positive'' audit outcome reflects full compliance, while a ``negative'' result necessitates immediate corrective actions and further evaluations. By embedding these auditing mechanisms, the DSA establishes a robust accountability framework to enhance the safety and transparency of digital platforms operating within the European Union \cite{RegulationEU20222022b}. This framework aligns with calls for accountability mechanisms in socio-technical systems that incorporate human oversight and actionable feedback loops  \cite{binns_fairness_2018}.

\subsection{Legal Requirements for Sampling in the Context of Systemic Risk Assessments and Independent Audits}

The DSA empowers the European Commission (EC) to provide additional rules through delegated regulations (DRs) (Art. 87), facilitating clarification and specificity in compliance processes, for example, the DR for researchers’ access to platform data \cite{DelegatedRegulationArt40}. The EC’s power has already been exercised to outline rules for conducting independent audits (Art. 37(7) DSA)\cite{DelegatedRegulationSupplementing2023}.
Auditing organizations must follow an audit procedure defined in Rec. 16 DR, which includes techniques such as data collection, methodological application, substantive analytical procedures, and other actions to gather and analyse evidence, excluding the issuance of opinions or reports. These methodologies should reflect the measures VLOPs and VLOSEs implement to mitigate risks (Art. 14 DR). Additionally, auditors are required to document and report any reasonable doubts arising from their assessment of internal controls and analytical procedures (Art. 13(2)(b) DR). The DR emphasizes the importance of robust methodologies, including qualitative and quantitative elements, in analysing systemic risks. For instance, Art. 3 DR delineates the scope of audits, requiring auditors to describe specific elements audited and the methodologies applied (Art. 37(4)(d) DSA; Art. 10 DR). Methodologies must link sampling to the minimization of detection risks (Art. 12(1) DR) and ensure alignment with the specific risks under investigation, such as the dissemination of illegal content (Art. 34(1)(a) DSA).

\textbf{Sampling as a Methodology in Audits} Sampling is explicitly recognized as a critical audit methodology to strengthen evidence (Art. 2(6) DR). Sampling methodologies are central to identifying risks that might otherwise remain obscured, particularly in algorithmically moderated environments where decisions are opaque \cite{geiger_operationalizing_2017}. Effective sampling ensures that a quantitative sample is representative of the content moderation quality of a platform, thereby enabling qualitative assessments of risks as outlined in Art. 34 DSA. Sampling methodologies must aim to minimize detection risks (Art. 12(1) DR; Art. 2 and Art. 3(2) (c) DR) and justify sample size and technique in audit and risk assessment reports (Art. 12(3) DR). The methodologies must account for materiality thresholds (Art. 10(2)(a) DR) and detail the analytical tests and procedures employed. Moreover, Art. 10(2)(b) DR demands the description of tests and other analytical procedures an auditing organization intends to use during their evaluation. According to Rec. 17 DR, a test is “an audit methodology consisting in measurements, experiments or other checks, including checks of algorithmic systems, through which the auditing organization assesses the audited provider’s compliance with the audited obligation or commitment”.

\textbf{Regional and Contextual Specificities } Art. 13 DR highlights the need for contextualized risk assessments, including regional and linguistic factors, particularly when analysing civic discourse and electoral debates (Art. 34(1)(c) DSA). For example, auditors must account for regional disinformation narratives, electoral laws, and political controversies \cite{marsdenRespondingDisinformationTen2020, kubler2021GermanFederal2023}. Art. 35(2) DSA further emphasizes the importance of presenting systemic risks by Member State and at the Union level. By considering these factors, auditors ensure that sampling methodologies are representative and relevant to the contexts being studied, aligning with Art. 13 DR’s mandate to address ``probability and severity of risks''.

\textbf{Methodological Requirements and Assurance} 
Auditors must describe how information sources were selected and assumptions tested with impacted groups (Art. 13(1)(a)(v) DR). Methodologies must address algorithmic systems, content moderation practices, Terms and Conditions enforcement, advertising, and data practices (Art. 34(2) DSA). Art. 10(2)(a) DR mandates that methodologies make sense of data both quantitatively and qualitatively, with substantive analytical procedures used to assess compliance risks (Rec. 18; Art. 10(2)(b) DR). Auditors must provide a ``reasonable level of assurance'' in their evaluation of a platform’s controls to mitigate identified risks (Art. 13(2) DR). Sampling methodologies, while critical, must be transparently documented in both risk assessment and audit reports to ensure methodological rigor and alignment with the DSA’s objectives. By emphasizing rigorous sampling and contextual specificity, the DSA and DR ensure that systemic risk assessments and independent audits are robust, representative, and capable of addressing the diverse challenges posed by digital platforms.

\section{MIXED METHODS – COMBINING STATISTICAL CONTENT SAMPLING TECHNIQUES AND CONTENT ANALYSIS FOR SYSTEMIC RISK ASSESSMENTS AND INDEPENDENT AUDITS}
This section addresses research question two: What sampling methods would be most relevant for internal risk assessments and independent audits? According to Webb and Wang, the aim of sampling is the representativeness of the sample itself that presents the “[. . . ] incidences selected accurately [and] portray[s] the population”  \cite[98]{webbTechniquesSamplingOnline2013}. In an ideal world, those analysing the sample would be working with a perfect ground truth sample that also includes information, for example, about the appropriate time frame for data collection or geographical dimension of data collection and context. However, such an ideal ground truth sample is usually not attainable \cite{zafaraniEvaluationGroundTruth2015}. This is not only because of the imperfect access to data from a third-party auditing perspective but also because platforms themselves with assumed ideal knowledge can have problems attaining a “perfect” ground truth sample. This is not only due to the subjectivity of the decision of what constitutes a violation of the law or TaC  \cite{kumarWatchYourLanguage2024} but also because of historic data bias, geolocation blocking, or protection from legal liability \cite{geigerPlatformLiabilityArticle2021}.

In the area of social media research, different methodological approaches exist on how to collect and analyse data. Data analysis within the context of systemic risk assessments and independent audits leads, according to Piña-García et al., to a knowledge gap regarding the data needed, the quantity of it, the questions and queries to ask, or the protection or quality of privacy  \cite{pina-garciaStandardSamplingMethodology2016}.
Sampling techniques used within the context of the DSA, however, can have several ways to create a sample that meaningfully represents the population and must take the individual systems, content, and platform design into account. According to Rougier, risk assessments can “be made with a varying degree of confidence, reflecting both the innate difficulty of the task, and the amount of resources available”  \cite[1081]{rougierConfidenceRiskAssessments2019}.

Online platforms samples can be drawn before publishing the content online (ex ante sample) or after the platform’s internal moderation process is in place (ex post sample). While the first content sample includes a realistic picture of the violations and content uploaded to platforms, the second content sample provides a snapshot of what users will see after content moderation has taken place. Furthermore, ex post content moderation samples do not include already filtered-out content. This content class is either deleted due to violations of the TaC or the law, or can be geo-blocked in certain areas. Both sampling time frames can provide meaningful information for assessing systemic risks and DSA compliance of platforms. While an ex ante sample might indicate what systemic risks are prevalent, trending, or novel, ex post content samples might provide further insight into internal content moderation quality, challenges, and areas for improvement. Ex ante sampling can also help identify potential algorithmic vulnerabilities before they result in significant harm, an approach advocated for in fairness auditing  \cite{raji_actionable_2019}.

Sampling which uses random chance is called probability sampling, in which repeated sampling using the same method results in different samples. Simple random sampling is the most intuitive method of random sampling: each member of a population has an equal chance to be included in the sample. Nonprobability sampling methods are deterministic in nature and always result in the same sample. The most common nonprobability sampling method is convenance sampling, where a researcher takes members of the population that are easiest to access without any use of random chance.
Simple random sampling can be used to create an auditable sample, for instance, to evaluate the prevalence of illegal content and TaC violations \cite{mubarakDetectingReasoningDeleted2023}. This method of sampling does not take “auxiliary information on the population” into account when selecting a sample \cite[12]{europeancommissionSurveySamplingReference2008}. Such a sampling method could be used for both ex ante and ex post samples to understand the overall content uploaded on platforms  \cite{RegulationEU20162016, pfefferThisSampleSeems2023, mubarakDetectingReasoningDeleted2023, morstatterSampleGoodEnough2013a}. This could contribute to understanding what violations are common on a platform. However, there are also downsides to random sampling techniques, for example, if unequal inclusion probabilities harm the representativeness of a sample  \cite{grafstromHowSelectRepresentative2014}. 

Ex post content moderation samples are more common in most research domains because of the data access and data collection possibilities. Researcher access depends on a platform’s willingness to provide an API and data access. To illustrate the use of a social media platform’s API, X (formerly Twitter) is a useful example, as social science research focused predominantly on the social media platform X, before free access to its API was terminated in 2023 \cite{pfefferThisSampleSeems2023}. 
Researchers usually choose a corpus that best fits to answer their research question\cite{andreottaAnalyzingSocialMedia2019}, for instance, by choosing specific hashtags that refer to the topics they are researching \cite{martiniBotNotComparing2021}. This approach, however, is open to biases and might not represent every aspect of a topic or conversation. Moreover, data returned by the API might be subject to inherent biases that are not detectable for external researchers \cite{rafailNonprobabilitySamplingTwitter}. 

When accessing the API of X, only  publicly available information (i.e. Tweets and metadata) can be accessed; a banned account, for instance, would not be included in a dataset \cite{pfefferThisSampleSeems2023}. However, the sampling strategy influences the quality of the data. To choose a representative sample, Rafail \cite{rafailNonprobabilitySamplingTwitter} proposes first defining a population pertinent to the investigation. That might be an “unbounded population”, which is best used with random sampling and includes every user and tweet. More often used in research are nonprobability sampling strategies that focus on “semi-bounded populations”, for which either the users or the content is chosen according to clear criteria pertaining to the research question. The users are chosen by consulting sources outside of the platform. Content restrictions are included by using hashtags. Finally, “bounded populations” focus on a sample of a very restricted user group, which focuses on a group of accounts. Ex post content moderation samples are limited as the sample is only collected as publicly accessible in that moment. 

However, Pfeiffer et al.  
\cite{pfefferThisSampleSeems2023} have shown that a large number of posts are removed either by platforms or users. For risk audits in the context of the DSA (Art .35(c)), deleted posts are important for platforms to show compliance with the provisions of risk mitigation by means of adapting content moderation systems.
In contrast to simple random sampling, the probability proportional to size sampling method takes information about the population of a sample into account to create an auxiliary variable that determines some aspect in relation to size and the population being studied. For creating representative samples with this technique, more accurate information about a population is needed to correctly estimate inclusion probability. In a scenario with perfect knowledge about platform violations, for example, the reported numbers of orders received according to Art. 9 and 10 DSA about illegal content in a specific Member State could be known, \cite{goodmanOnlineCommentModeration} which might facilitate the  estimating of good measurements for probability sampling to create a representative sample of auditable platform content. Following this logic, the prevalence of orders that refer to violations of the law could be used as a criterion to interpret samples’ underrepresentation or overrepresentation of problematic legal content. However, it is highly likely that even the platforms themselves do not have perfect knowledge about themselves in that sense, for example, in cases of terrorist content and deletion obligations which link platforms’ liability starting only after they receive removal orders per Art. 3 \cite{RegulationEU20212021} as reflected by the non-existence of a general  monitoring obligation of platforms for their uploaded content according to Art. 8 DSA.

When a population has substantially different subpopulations called strata, sampling within each stratum independently may have statistical and practical advantages. Stratified sampling allows different sampling strategies within strata. Stratified sampling in comparison to the other sampling techniques presented makes use of information about a population, but also considers the specificities of a sample by dividing the population into separate strata  \cite[12]{europeancommissionSurveySamplingReference2008}. According to Bilon and Clemente, this approach could start from “the date and time the post is created, content type, message, link, and post story” for which “content type” is the simplest category to base the strata on  \cite[52]{bilonEvaluationSamplingMethods2020a}.
These divided groups are called strata and are “non-overlapping subpopulations”  \cite[12]{europeancommissionSurveySamplingReference2008}. The creation of such groups, also called strata, can be based on additional information about the data, such as content type, and can increase precision   \cite{bilonEvaluationSamplingMethods2020a}. 
Therefore, stratified sampling allows different sampling needs for subpopulations and can be adjusted, for example, to specific systemic risks or content types. In the case of systemic risks such as the groups most impacted in line with Art. 13 DR, regional aspects of content according to Art. 13(2) DR or correlating languages of Member States might be useful categories for defining strata. Content types, on the other hand, could include the selection of images, video content, or comments to be included in a sample that addressed a specific risk, for example, gender-based violence. Grafström and Schelin highlight the limitations of stratified sampling as a method. According to them, the classes would get too numerous or small to be able to meaningfully interpret the results  \cite{grafstromHowSelectRepresentative2014}. Additionally, the sampling time is highlighted in the context of content samples on Facebook according to Bilon and Clemente for posts “as elementary unit” under days, weeks, and simple random sampling (with and without replacement)  \cite[44]{bilonEvaluationSamplingMethods2020a}.

\subsection{On the alignment of sampling techniques and systemic risks}
Because systemic risks are different in nature, aligning them to sampling methods is a critical part of preparing audit methodologies. One example of sampling criteria involves testing compliance with Art. 23 DSA, which sets rules for handling users who repeatedly exhibit malicious behaviour on platforms. Sampling can focus on user accounts, including those not classified as repeatedly violating Art. 23 DSA and those that have exhibited such behaviour  \cite{zannettouWhoLetTrolls2019}. User IDs or account handles can serve as legitimate sampling criteria to construct an auditable sample representing, for example, the risk of accounts spreading misinformation.

\textbf{The challenge of Member State-specific risk } Another important aspect of sampling involves Member State- specific risk domains, which requires sampling based on geographic location (e.g. an uploader’s location or a recipient’s location if no VPN is used), content targeting specific Member States, or content language\cite{kubler2021GermanFederal2023}. Evaluation of content can have many layers, for instance, a TikTok post might include text in German in the background, spoken French by the creator, English and French hashtags, and a description in English [see TikTok language and event-based]. Comments can vary in language, and emojis might hold different meanings depending on the language context  \cite{aliexpress_digital_2024}. These linguistic complexities affect keyword selection, abbreviations, and acronyms, which are critical for creating meaningful samples for compliance testing under the DSA.

\textbf{The challenge of keyword/hashtag selection } Hashtags, while sharing challenges with keywords, provide additional layers of content metadata and can inform event-based testing \cite{tiktok_tiktok_2024} and moderation decisions \cite{aliUnderstandingEffectDeplatforming2021}. Similarly, sampling based on content similarities can enhance audits in line with content hashing initiatives  \cite{leeDetectingChildSexual2020, gerrardHashtagCircumventingContent2018, gorwaAlgorithmicContentModeration2020, westlakeComparingMethodsDetecting2012, baurMethodsEmpiricalSocial2022}.

Sampling based on meaningful time frames is another challenge [ref table sampling intervals]. For instance, auditing systemic risks affecting electoral processes (Art. 34(1)(d) DSA) requires ensuring data collection coincides with election periods in Member States \cite{tiktok_tiktok_2024}. Missing relevant time frames could result in underrepresentation of violations in a sample and increase detection risks, undermining an audit’s accuracy and relevance.

\textbf{The challenge of capturing context } Content specificity must also inform sampling decisions in line with Art. 12 and 13 of the DR. Content samples may include main posts, threads, sub-threads, comments, likes, or warning labels \ref{comparison_all}. Auditors may require additional contextual information, such as chronological timelines, to draw meaningful conclusions for specific risk categories. For example, a seemingly neutral post like “yes, let’s do that” might reveal a violation when linked to prior antisemitic statements. Conversely, counter-speech, which may appear violative in isolation, could only be understood as compliant in context \cite{wagner_mapping_2024}. Chronological and contextual analysis is thus essential for assessing compliance meaningfully within digital conversations.

\textbf{The challenge of sampling interval selection } The feasibility of analysing sample sizes also varies with con- tent type. For example, YouTube’s diverse content – ranging from Shorts (brief videos) to long-form videos and live streams – creates challenges in determining appropriate sample sizes  \cite{husseinMeasuringMisinformationVideo2020}. Feasibility considerations must be justified under Art. 12(3) DR.

\textbf{The challenge of violation location } Additionally, the positioning of violations within content varies across platforms. On X (formerly Twitter), violations could appear in the main post, images, text within images, hashtags, or comments. These variations, coupled with platform-specific designs [e.g. Pinterest’s pins], algorithms, and features, add complexity to the creation of representative sampling methodologies. Lovato et al., for example, compare the concept of compliant user consent across platforms  \cite{lovato_limits_2022}. Auditors must consider these factors to develop auditable and representative samples and explain their rationale.
These examples underscore the complexity of selecting and applying sampling methods for compliance testing under the DSA. We propose using (legal) content analysis as a substantive analytical procedure for auditing VLOPs’ and VLOSEs’ compliance (e.g., \cite{snapchat_snap_2024, google_report_2024, google_report_2024-1, google_report_2024-2, google_report_2024-3, google_report_2024-4}). This approach can provide valuable insights into systemic risks identified in Art. 34 DSA  \cite{kubler2021GermanFederal2023}.

\subsection{Legal Content Analysis}

Legal content analysis adapts methods and traditions from disciplines such as political science and sociology  \cite{bar-zivContentAnalysisApproach2021}. The emphasis on legal content analysis aligns with Bogiatzis-Gibbons’ advocacy for reclaiming democratic oversight in AI systems for the purpose of achieving oversight of content moderation systems. This approach can guide the development of codebooks and coding practices that align with democratic values and legislative requirements under the DSA  \cite{bogiatzis-gibbons_beyond_2024}. Through this approach, empirical legal scholars aim to bring the rigor of social science to the study of law \cite[1]{brookPoliticsCodingSystematic2021}. Brook posits that content analysis bridges traditional legal methodologies, such as doctrinal research, with quantitative empirical elements  \cite[3]{brookPoliticsCodingSystematic2021}. Hall and Wright highlight its value in increasing objectivity within legal research\cite{hallSystematicContentAnalysis2008}. Content analysis involves emergent coding, in which codes arise from data itself (e.g. analysing reasons for violations in a sample), or a priori coding, which uses predefined criteria such as a codebook\cite{salehijamValueSystematicContent2018}. Hall and Wright  further discuss
the iterative process of creating and refining codebooks to categorize data\cite{hallSystematicContentAnalysis2008}, with testing and adjustments ensuring clarity and usability\cite{guestHandbookTeamBasedQualitative2007}. 
%For DSA audits, such a codebook could incorporate Member State legal norms and Terms and Conditions (TaC) categories tied to content moderation actions.

Bar-Ziv outlines three stages of content analysis: (1) Data making, (2) Inferring, and (3) Narrating  \cite[475]{bar-zivContentAnalysisApproach2021}. Salehijam, however, emphasizes formulating a research question as the initial step, followed by data collection, coding, reasoning, and, finally, presenting findings to the target audience \cite{salehijamValueSystematicContent2018}. Hall and Wright argue that content analysis enables legal scholars to derive qualitative insights and, subsequently, quantify these for deeper legal analysis, as needed for DSA audits  \cite{hallSystematicContentAnalysis2008}. Systematic application of content analysis is often referred to as Systematic Content Analysis (SCA), a replicable technique for analysing texts, including case law and legislation  \cite[35]{salehijamValueSystematicContent2018}.

While effective, content analysis has limitations. Brook cautions against the lack of comparability across fragmented legal domains \cite[1]{brookPoliticsCodingSystematic2021}, an issue relevant to European Member State law and the DSA. Bar-Ziv highlights the problem of missing data, such as cases resolved outside courts  \cite{bar-zivContentAnalysisApproach2021}. For DSA-related audits, analogous issues arise when deleted content or risks are excluded from samples, potentially skewing metrics and underrepresenting risks. Ensuring robustness in content analysis involves coding samples by multiple individuals to compare subjective decisions, with inter-coder reliability measured using metrics like Krippendorff’s alpha or Cohen’s kappa  \cite{hsuInterraterAgreementMeasures2003}.

For systemic risk assessments and external audits, we propose a priori coding with risk-specific codebooks tailored to the risks being assessed. For example, codebooks should include legal provisions linked to deletion or geo-blocking obligations under Art.34(1) (a) DSA for risks such as disseminating illegal content. Non-legal risks, like negative effects on electoral processes, require bespoke coding schemes  \cite{kubler2021GermanFederal2023}. After developing risk-specific codebooks, content analysis should involve double-coding each sample by at least two coders. Measuring inter-coder reliability ensures transparency and robustness in subjective decisions during the auditing process.
We further propose a ``Moderation Decision Supervisory Team'', comprising experts on the analysed risks, to resolve disagreements between coding teams. For instance, if legal coding team 1 does not classify a video as violating the TaC, but legal coding team 2 does, the supervisory team would make the final determination. This structure enhances the quality and reliability of audits by enabling external auditors to generate both qualitative and quantitative insights into the content moderation practices of VLOPs and VLOSEs, aligning with Art. 10 DR.

\section{ENSURING REPRESENTATIVENESS OF CONTENT SAMPLES}

In general, representativeness ensures that the sample reflects the population characteristics on all the variables of interest \cite[278]{grafstromHowSelectRepresentative2014}. According to Art. 12(2) DR, both the sample size and the methodology used to assess it must be representative. However, as discussed in earlier sections, sampling methods and required sample sizes vary based on platform, content type, risk assessed, content distribution, assessment period, and the availability of historical or real-time data. These diverse requirements underscore the complexity of creating representative samples for auditing. Notably, the selection of sample size and methodology often depends on the subjective judgment of auditors  \cite[1084–1085]{rougierConfidenceRiskAssessments2019} and may be influenced by differing views on “how to do audits right”.

As highlighted by Pfeffer et al., the quality of sampling and methodology can decline if data  are not collected continuously, leading to temporal bias in the sample. Their findings indicate that data availability decreases by approximately 2\% per month within the first 18 months after content publication \cite[6]{pfefferThisSampleSeems2023}. Similarly, the choice of API for data sourcing (e.g. Streaming API vs Academic Research API) can result in varying data samples  \cite{alkulaibCollectEthicallyReduce2020}.

This decay in data availability starts almost immediately. For example, in one instance, “1 out of 30 Tweets collected 24 hours after posting was unavailable on the Academic API 10 seconds after posting” \cite{pfefferThisSampleSeems2023}[49:7]. Additional challenges include representativeness issues arising from distributed data centres, differing geo-blocking policies, and accessibility conditions. Under such circumstances, achieving a highly representative sample, let alone a perfect ground truth sample, becomes increasingly difficult.
Pfeffer et al. also highlight the challenges posed by deplatforming user accounts. For example, when U.S. President Trump’s Twitter account was deplatformed, metadata associated with the account (e.g. retweets) was also rendered unavailable for analysis \cite{PermanentSuspensionRealDonaldTrump}. Such scenarios align with risks identified under Art. 34 DSA, including malicious user behaviour and the frequent provision of manifestly illegal content per Art. 23 DSA. While larger sample sizes may mitigate some sampling method challenges via the law of large numbers, the methodology itself significantly influences sample quality and, consequently, representativeness \cite{malikFrameworkCollectingYouTube2017}. For example, Siersdorfer et al. conducted an in-depth study of commenting and rating behaviour using a sample of over six million comments on 67,000 YouTube videos, utilizing metadata such as titles, tags, categories, descriptions, upload dates, and user engagement statistics\cite{siersdorferHowUsefulAre2010a}. Moreover, qualitative analysis can meaningfully enhance representativeness \cite{de-lima-santosInstagrammableDataUsing2021}.

Given these challenges, creating a representative sample under the DSA requires detailed documentation and robust arguments tailored to individual audits. There is no universal definition of sample representativeness; instead, auditors and platforms must perform context- and risk-dependent analyses that are both visible and transparent \cite{eyert_rethinking_2023, flyverbomDigitalAgeTransparency2016}. This aligns with Art. 12(3) DR’s requirement to justify the selection of sample size and methodology, as well as Art. 13 DR’s requirement to describe the sources of information for audit evidence (Art. 2(6) DR). By providing thorough documentation on risk estimation, distribution, error, and population, auditors can foster a better understanding of subjective decisions, increase the robustness of their studies, and improve the reproducibility of the auditing process. Such explanations also contribute to defining the materiality threshold in accordance with Art. 37(4) DSA(d) and Art. 10(2)(a) DR.

\section{CONDUCTING INDEPENDENT AUDITS AND SYSTEMIC RISK ASSESSMENTS}

According to the DR, the goal of sampling is to minimize the detection risk of the audit (Art. 2(3) DR and Art. 12(1) DR). Detection risk is defined in Art. 2(8) DR as ``the risk that the auditing organisation does not detect a misstatement that is relevant for the assessment of the audited provider’s compliance with an audited obligation or commitment''. Thus, auditing risk refers to instances when external auditors fail to identify mistakes, omissions, or misunderstandings during their audit. The DR also defines two additional risk categories: inherent risk and control risk. The role of auditors and the element of subjectivity in deciding in audit methodologies and certainty levels however is critiqued under the formulation of the DSA \cite{terzis_law_2024}.

Inherent risk refers to the vulnerabilities stemming from ``the nature, the design, the activity and the use of the audited service, as well as the context in which it is operated, and the risk of non-compliance related to the nature of the audited obligation or commitment'' (Art. 2(1) DR). This highlights the importance of individual designs and services offered by a VLOP or VLOSE. For example, the presence of malicious users may vary based on whether the platform features a marketplace or live-streaming functionality. Control risk, on the other hand, pertains to the possibility of the audit being misdirected or unable to identify risks accurately, which cannot be resolved through internal controls (Art. 2(11) DR). For instance, external auditors might fail to identify the risk of gender-based violence affecting groups like LGBTQ+ users.
Under Art. 37(2) DSA, VLOPs and VLOSEs must provide ``cooperation and assistance necessary to enable [the auditors] to conduct those audits in an effective, efficient and timely manner, including by giving them access to all relevant data and premises and by answering oral or written questions''. If an auditing organization lacks access to necessary data and cannot adequately analyse, identify, and assess risks (Art. 13 DR), it must note these shortcomings
in the audit report (Art. 37(5) DSA). However, such data inaccessibility can increase detection risk and undermine the robustness of the analysis.
This definition underscores the critical role of context in risk classification and evaluation. Context is also essential for conducting systemic risk assessments and external audits. Transparency reports (Art. 15, 24, and 42 DSA) and Statements of Reason (Art. 17 DSA) benefit from contextualization and can serve as additional evidence in audits. Art. 13 DR mandates that audits consider the probability and severity of risks, aligning these factors with the appropriateness of the assessment and the minimization of detection risk. Additionally, the probability and severity of risks must be assessed for the groups most impacted. For example, the LGBTQ+ community might face a heightened risk of gender-based violence (Art. 34(1)(d) DSA), or female politicians may be targeted in disinformation campaigns before elections 
\cite{leeMeasuringDiscriminationLGBTQ2017, kubler2021GermanFederal2023}.

These considerations add complexity to the auditing process, which may also be constrained by available resources. Regional and linguistic aspects of detection and control risks must also be addressed, documented, and justified within audit reports. Member State-specific samples should be collected and analysed in their respective languages to account for these regional and linguistic differences. Additionally, risks in these samples must be tested on groups most impacted, with documentation explaining who these groups are and why they were chosen. Such transparency and rigor are essential for maintaining the credibility of the audit process.
To minimize detection risk, multiple samples should be drawn within representative time frames, such as the period preceding an election. These samples should cover all Member States and their languages, addressing regional and linguistic aspects comprehensively. The auditors’ decisions must be documented and justified in the audit report. If an audit lacks sufficient rigor and a reasonable level of assurance regarding systemic risks under the DSA, the credibility of the audit report and process may suffer. Transparent methodologies, including clear documentation of sampling choices, are essential for maintaining the integrity of audits  \cite[p. 429]{raji_actionable_2019}. This could lead to a decrease in the reasonable level of assurance and negatively impact detection and control risks.

\section{COMPARISON OF SYSTEMIC RISK ASSESSMENTS AND AUDIT REPORTS}
\label{comparison_all}

In this section we compare the first systemic risk assessment reports of VLOPs and VLOSEs and the first audit reports, analysing how these reports have addressed sampling as a methodology to assess systemic risks and to minimize detection risk. We analyzed 19 reports from 2023, observing significant variations in the depth of information provided, methodological rigor, and transparency across platforms and auditing firms. The analysis focuses on the strengths and weaknesses of the sampling methodologies, the consistency of data across sources, and the implications for evaluating compliance the DSA and the DR. In the following sections we compare first systemic risk assessment reports, and second audit reports, with a focus on the following: sampling as a method used, the time frame, or the content used in sampling. 

\subsection{Systemic risk assessment reports}

Within systemic risk assessment reports, platforms vary in the level of detail they provide regarding methodologies (see \ref{tab:platform-sampling}). The best practice example is Snapchat, which defines confidence intervals, a  ±5\% margin of error, and employs the KPI ``Policy Violating Prevalence'' (PVP) rate, encompassing Terms and Conditions (TaC) violations and illegal content  \cite[p. 5, p. 78, p. 216.]{snapchat_snap_2024}. Their inspections are based on daily random sampling, aligning with proactive risk mitigation strategies, such as monitoring Public Stories. YouTube, by contrast, uses the ``Violative Video Rate'' (VVR) as a random sampling metric to identify gaps in automated systems.  This metric estimates the percent of undetected violations on the platform,
evaluated by humans using ``statistically significant samples'' \cite[p. 100]{google_report_2024-4}. However, this approach lacks clarity regarding the sample size, the method used to calculate it, and details about the sampled content (e.g. comments, YouTube Shorts, or videos, including min./max./average length). Despite these gaps, the mixed-methods approach in the VVR aligns with the methodology proposed in this article.
LinkedIn employs ``positive samples'' (assumed to mean false positives) of accounts and content on a weekly or monthly basis to evaluate the precision of automated content moderation using ``regular sampling'' but does not specify sample size or confidence intervals  \cite[p. 19, p. 23, p. 28.]{linkedin_linkedin_2024}. Similarly, AliExpress conducts weekly quality checks on product listings, focusing on multilingual content in high-risk product categories, both internally and with external professional service providers \cite[p. 24, p. 31.]{aliexpress_digital_2024}.

\begin{table*}[ht]
\scriptsize 
\caption{Platform Sampling Methods and Analysis in Relation to Systemic Risks}
\label{tab:platform-sampling}
\begin{tabular}{p{1.5cm} p{1.5cm} p{2.5cm} p{1.5cm} p{1.5cm} p{1cm}}
\toprule
\textbf{Platform} & \textbf{Sampling Method} & \textbf{Method of Analysis} & \textbf{Frequency} & \textbf{Focus Area} & \textbf{Source} \\
\midrule
AliExpress         & Not clear               & Multilingual sampling for illegal listings                            & Weekly               & Illegal listings compliance       & pp. 24–31 \\
Amazon Inc.        & Proactive sampling      & Manual and automated compliance controls                              & Ongoing              & Seller verification, safety       & pp. 6–18 \\
Apple              & Not clear               & Manual triage integrated with automated systems                       & Annual               & App Store risks                   & pp. 66–81 \\
Booking.com        & Not clear               & Scenario-based systemic risk assessment                               & 5 months             & Reviews and discrimination        & pp. 1–2 \\
Google Platforms   & Statistical sampling    & Random and flagged content review                                     & Periodic             & Content policy compliance         & p. 100 \\
LinkedIn           & Positive sampling       & QA audits and prevalence exercises                                    & Weekly, Monthly      & Abuse trends, algorithm QA        & pp. 19, 28 \\
Facebook           & Not clear               & Governance-based feedback loops                                       & Continuous           & Moderation governance             & p. 59 \\
Instagram          & Not clear               & Governance measures and user feedback                                 & Continuous           & Harmful content moderation        & p. 59 \\
Microsoft Bing     & Not clear               & QA sampling and user activity evaluation                              & Annual               & Search and ad safety              & pp. 3–22 \\
Pinterest          & Not clear               & Moderation error trend analysis                                       & Ongoing              & Moderation error reduction        & p. 7 \\
Snapchat           & Random sampling         & Testing for policy violation prevalence rates                         & Daily                & Detection blind spots             & pp. 5–216 \\
TikTok             & Targeted sampling       & Context-specific reviews for election content                         & Event-driven         & Election misinformation           & p. 33 \\
Wikimedia          & Not clear               & Human rights-based approach                                           & Periodic             & Risk register                     & Online resource \\
X (Twitter)        & Not clear               & Crisis management and systemic risk profiling                         & Annual               & Crisis management, risks          & pp. 6–10 \\
Zalando            & Not clear               & Partner business compliance reviews                                   & Annual               & Partner content compliance        & pp. 1–30 \\
\bottomrule
\end{tabular}
\end{table*}

Pinterest incorporates ``calibrated sampling'' for ``Root Cause Analysis'' and ``Error Trend Analysis'' as part of its quality assurance program \cite[p. 7p. 7]{pinterest_digital_2024}. Meta, on the other hand, provides limited insight into sampling methodologies, mentioning only ``continuous testing'' without elaborating further  \cite[p. 59 Meta]{meta_systemic_2024}. TikTok employs ``scenario-specific sampling'' to detect election-related risks, such as misinformation, focusing on top-rated videos and high-priority events \cite[p. 33]{tiktok_tiktok_2024}.
Regarding sampling frequency, platforms demonstrate varied approaches: continuous (Snapchat, Meta), periodic (YouTube, Pinterest), event-driven (TikTok), or specified intervals (LinkedIn, AliExpress)

%\paragraph{In sum, platforms utilize varied and context-specific sampling methodologies. Platform- specific challenges in compliance and risk mitigation, such as those involving Facebook’s ad delivery systems, highlight the importance of tailored sampling and testing methodologies \cite[p. 15]{ali_discrimination_2019}. Such include continuous (e.g. Snapchat, Meta), periodic (e.g. YouTube, Pinterest), event-driven (e.g. TikTok), and interval-based approaches (e.g. LinkedIn, AliExpress), with significant differences in transparency, sampling precision (e.g. defined confidence intervals, error margins), and the integration of qualitative and quantitative metrics to assess systemic risks and optimize automated content moderation systems.}

\begin{table*}[ht]
\scriptsize % Reduce font size for compactness
\caption{Comparative Analysis of Platform Strengths and Weaknesses}
\label{tab:platform-strengths-weaknesses}
\begin{tabular}{p{2cm} p{5cm} p{5cm}}
\toprule
\textbf{Platform} & \textbf{Strengths} & \textbf{Weaknesses} \\
\midrule
AliExpress         & Multilingual checks; high-risk targeting             & Limited transparency on emerging risks \\
Amazon Inc.        & Automation and manual reviews; scalable accuracy     & Unclear handling of false positives \\
Apple              & DSA-aligned; stakeholder consultation                & Low intervention thresholds \\
Booking.com        & Tailored transaction-specific approach               & Narrow scope beyond reviews \\
Google Platforms   & Wide sampling and broad coverage                     & Minimal systemic risk insights \\
LinkedIn           & QA audits; abuse pattern refinement                  & Lacks detailed long-term risk plans \\
Facebook           & Appeals-driven feedback loops                        & Lacks broader systemic context \\
Instagram          & Feedback integration                                 & Weak systemic abuse data focus \\
Microsoft Bing     & Combines ranking with QA                             & Limited external transparency \\
Pinterest          & Trend analysis with consistent QA                    & Minimal focus on systemic risks \\
Snapchat           & Statistically robust (95\% CI)                       & Gaps in emerging risk handling \\
TikTok             & Context-specific targeting                           & Lacks general risk evaluation \\
Wikimedia          & Human rights-centered                                & Lacks sampling/selection criteria details \\
X (Twitter)        & Focus on hate speech/misinformation                  & No granular systemic data \\
Zalando            & Reviews of recommender systems                       & No external validation \\
\bottomrule
\end{tabular}
\end{table*}

\subsection{Audit reports}

Audit reports conducted by auditing firms provide critical insights into methodologies, as required by Arts. 12 and 13 of the DR as illustrated in Table \ref{table_audit_reports}. The sampling methods across the referenced audit reports exhibit notable variations in their complexity, scalability, and applicability to different compliance contexts.  A notable example is Snapchat’s audit report, which clearly specifies a materiality threshold of  95\% and an actual or projected error margin of 5\% \cite[p. 63]{ey_independent_2024-8}. Similarly, all audit reports conducted by EY include dedicated sections on sampling methods  \cite{ey_independent_2024, ey_independent_2024-1, ey_independent_2024-2, ey_independent_2024-3, ey_independent_2024-4, ey_independent_2024-5, ey_independent_2024-6, ey_independent_2024-7, ey_independent_2024-8, ey_independent_2024-9}. These reports detail sampling approaches, such as the selection of sample sizes: ``e.g., a sample of 25 when the population is greater than 250 occurrences or 10\% of the population size, with a minimum sample of 5 when the population is less than 250 occurrences''  \cite[p. 11]{ey_independent_2024-1}.

\begin{table*}[t]
\centering
\scriptsize 
\caption{Platform Sampling Methods and Analysis in Relation to Audit Reports}
\resizebox{\textwidth}{!}{%
\begin{tabular}{p{2cm}p{2cm}p{3cm}p{4cm}p{2.5cm}}
\toprule
Category & Auditor & Sampling Method & Context Information & Source \\
\midrule
AliExpress & KPMG & Random sampling with risk-adjusted selection size. & Random selection considers illegal exceptions for >5\% of the selection, linking population size to significance levels. & AliExpress DSA Assurance Report, KPMG, Sept. 2024, pages 12, 35, 75, 98-99, 108. \\
Amazon Inc. & KPMG & Random sampling with risk-adjusted selection size. & Population size linked to significance levels. & Amazon DSA Audit Report 2024, pages 34, 95, 98-100. \\
Apple & EY & Attribute sampling techniques. & Sample size based on population size, with a focus on valid/invalid attributes. & Apple DSA Audit Report 2024, pages 16-17, 19, 30-31, 38-49, 51-62, 89-96, 100, 115, 125. \\
Booking.com & Deloitte & Statistical and nonstatistical methods. & Sample sizes determined by population size, control failure risk, and tolerable deviation rate. & Booking.com DSA Audit Report 2024, pages 14, 82, 90, 95, 105, 108, 113, 137, 145. \\
Google Platforms & EY & Attribute sampling techniques. & Sample size ensures appropriate attribute representation. & Google DSA Audit Reports 2024, pages 13-14, 27, 38-43, 45, 47, 49, 51, 55-57, 66, 75, 78, 80, 96-97. \\
LinkedIn & Deloitte & Statistical and nonstatistical methods. & Sample sizes based on population, control risk, and tolerable deviation rate. & LinkedIn DSA Report 2024, pages 7, 17-18, 30, 32, 38, 41, 43-45, 49, 51, 53-54, 56-57, 58, 61-62, 64, 72-73, 75, 84, 86, 94-95, 100-101, 103, 106, 135, 137. \\
Facebook & EY & Attribute sampling techniques. & Minimum sample size established for tested attributes. & Facebook DSA Audit Report 2024, pages 11-16, 18, 25-34, 36-37, 45-52, 56, 59-61, 80. \\
Instagram & EY & Attribute sampling techniques. & Same methodology as Facebook, targeting attributes. & Instagram DSA Report 2024, pages 11-16, 18, 25-35, 37-38, 46-51, 53, 57, 60-62, 79, 81. \\
Microsoft Bing & Deloitte & Statistical and nonstatistical methods. & Sample sizes based on population and control risks. & Microsoft Bing DSA Audit Report 2024, pages 8, 32, 107. \\
Pinterest & EY & Sampling aligned with regulations, considering risk evaluations. & Challenges included limits in algorithm assessments. & Pinterest-2024-DSA-EY-Independent-Audit-Report, pages 6, 12, 19, 25. \\
Snapchat & EY & Statistical sampling focusing on transparency metrics. & Reviewed transparency in ads, moderation, and personalization systems. & Snapchat DSA Report 2024, pages 12, 66, 80, 104. \\
TikTok & KPMG & Random sampling with size linked to control risk. & Population-linked significance testing. & TikTok DSA Audit Report 2024, pages 9-12, 21, 25, 28, 36, 38, 41-43, 49-50, 52, 59, 62, 68, 78, 86, 91-92, 96, 99-100, 105. \\
Wikimedia & Holistic AI & Non-statistical methods. & No statistical data provided. & Wikipedia DSA Audit Report 2024, pages 104, 107. \\
X (Twitter) & FTI Consulting & Random sampling. & Randomized communication tests for clarity and consistency. & TIUC DSA Audit Report 2024, pages 55, 65, 100, 113, 139-141, 153, 199-200, 207, 261, 319. \\
Zalando & Deloitte & Random sampling. & Cross-verification using legal product databases. & Zalando DSA Audit Report 2024, pages 20, 22, 34, 36-39, 41-42, 47, 49, 51-52, 69-70, 82, 85, 87-88, 96, 119, 121, 160, 164, 167, 198. \\
\bottomrule
\end{tabular}%
}
\label{table_audit_reports}
\end{table*}

EY’s reports consistently employ ``attribute sampling to determine if the sample selected has the desired attribute (e.g. correct or incorrect, present or absent, valid or not valid),'' forming a ``qualitative statistical [random] sample''  \cite[p. 13]{ey_independent_2024-6}. These methods evaluate whether specific attributes (e.g., presence, correctness, or validity) are consistently upheld within the sample population. The reports also highlight that no other sampling approaches are applicable when the populations lack quantitative dimensions, such as monetary balances in financial audits \cite[p. 13]{ey_independent_2024-6}. In cases of exceptions within the sample selection, EY addresses the need to extend sample sizes to create substantive evidence, noting, ``e.g., for sample sizes of 25, we tested at least 15 additional items or 40 in total'' \cite{ey_independent_2024, ey_independent_2024-1, ey_independent_2024-2, ey_independent_2024-3, ey_independent_2024-4, ey_independent_2024-5, ey_independent_2024-6, ey_independent_2024-7, ey_independent_2024-8, ey_independent_2024-9}. However, its application is limited in highly dynamic systems where attributes may evolve over time, potentially leading to outdated evaluations. Furthermore, the reports lack details on what those attributes are and which audit criterion or systemic risk category they link. 

In contrast, \textit{KPMG} audits do not provide the same level of detail regarding sampling methodologies in their reports \cite[p. 8]{deloitte_microsoft_2024}. Deloitte, on the other hand, employs both statistical and non-statistical sampling methods  and do not provide enough detail on understanding detailed decisions on samples, content types, sampling frequency, time frames, or confidence intervals.. However, reports audited by \textit{KPMG} (e.g., TikTok, AliExpress, and Amazon Inc.) predominantly employ random sampling with increasing selection size linked to risk levels associated with tested controls. \textit{Deloitte}, (e.g., LinkedIn and Zalando)  on the other hand, employs both statistical and non-statistical sampling methods (leveraging auditor judgment to define tolerable error rates and sample sizes)  \cite[p. 8]{deloitte_microsoft_2024}. However, \textit{Deloitte} reports only indicate the use of mathematical models for determining sample sizes, without specifying which models were used or their application to auditing criteria, risk categories, or content types. Nevertheless, it introduces subjectivity, which may impact reproducibility and comparability across audits. When auditors decide not to use statistical sampling is also not indicated in the audit reports, which makes detailed evaluation inaccessible.

Certain reports offer more granular details about sampling practices. For example, Zalando’s audit report describes the testing of a sample of 16 notices, with a benchmark time frame for reply set at 30 days  \cite[p 37]{deloitte_independent_2024-1}. Such variations in reporting practices underscore the need for standardization in audit methodologies to enhance transparency and comparability across platforms. 

%\paragraph{Audit reports reveal significant variations in sampling methodologies across firms, with EY providing detailed accounts of sampling approaches (e.g. attribute sampling, qualitative statistical sampling) and clearly defined thresholds (e.g. materiality threshold of  95\%, error margin of 5\%), while firms like Deloitte while firms like Deloitte and KPMG lack comparable transparency, with Deloitte noting only the use of unspecified mathematical models and KPMG offering limited methodological insights; these disparities highlight the critical need for standardized auditing practices to ensure consistent, transparent, and comparable assessments of systemic risks.}

\section{CONCLUSION AND FURTHER RESEARCH}
This article examined the strengths and weaknesses of quantitative and qualitative analytical methodologies in systemic risk assessments and audits under the DSA and DR, emphasizing the critical role of sampling as a methodological foundation. By analyzing 19 reports from VLOPs, VLOSEs, and independent auditors, we highlighted significant disparities in the transparency, robustness, and applicability of sampling techniques. These findings underscore the need for standardized yet adaptable approaches to address the complexities of socio-technical systems.

Our analysis identified systemic risk assessment reports as generally offering platform-specific metrics tailored to unique operational risks. For example, Snapchat's ``Policy Violating Prevalence'' and YouTube's ``Violative Video Rate'' illustrate how statistically robust sampling can illuminate systemic risks and guide proactive mitigation strategies. However, inconsistent documentation of sample sizes, criteria, and methodologies limits the replicability and transparency of many reports. Conversely, audit reports often focus on compliance and scalability,  providing detailed sampling thresholds, while other firms lack comparable methodological transparency. This variability highlights a gap in aligning systemic risk assessment and audit methodologies.

We proposed a mixed-method approach that integrates quantitative sampling with qualitative legal content analysis, balancing statistical rigor and contextual depth. Sampling methodologies should be tailored to the platform's characteristics, risks, and context, ensuring representativeness and relevance. Qualitative legal content analysis complements these methods by contextualizing violations and aligning findings with systemic risk categories outlined in the DSA.

The mixed-method approach offers actionable insights for both platforms and regulators. First, it enables auditors to identify and address gaps in content moderation systems, linking systemic risks to measurable KPIs. Second, it provides platforms with a framework for adapting sampling methodologies to dynamic risks, such as those related to misinformation, election integrity, or gender-based violence. Lastly, the approach ensures that audits not only verify compliance but also contribute to the continuous improvement of platform governance.

By bridging methodological gaps and addressing variability in current practices, this article contributes to the growing field of DSA compliance research. Our proposed mixed-method framework offers a pathway for advancing fairness, accountability, and transparency in the governance of socio-technical systems.

\bibliographystyle{ACM-Reference-Format}
\bibliography{Main}

%%Start the appendix with the ``\verb|appendix|'' command:
%%\begin{verbatim}

%%\end{verbatim}
%%and note that in the appendix, sections are lettered, not
%%numbered. This document has two appendices, demonstrating the section
%%and subsection identification method.

%%\begin{acks}
%%To Robert, for the bagels and explaining CMYK and color spaces.
%%\end{acks}


%%\appendix

%%\section{Research Methods}

%%\subsection{Part One}

\end{document}